\documentclass{JHEP3}
\usepackage{amsmath}

\newcommand{\baa}{\begin{eqnarray}}
\newcommand{\ea}{\end{eqnarray}}
\newcommand{\pv}[1]{{-  \hspace {-4.0mm} #1}}

\newcommand{\h}{Hamiltonian}

\def\nn{\nonumber \\}
\def\r{\rho}
\def\rn{\rho_0}
\def\m{m_0}
\def\te{\tilde{\eta}}
\def\e{\eta}
\def\l{\lambda}
\def\p{\partial}
\def\o{\omega}
\def\O{\Omega}
\def\td{\tilde{d}}
\def\tc{\tilde{c}}

\title{Solitons and excitations in the duality-based  matrix model}

\author{Ivan Andri\'c, Larisa Jonke, Danijel Jurman \\  
Theoretical Physics Division, Rudjer Bo\v skovi\'c Institute, \\
POB 180, 10000 Zagreb, Croatia \\
E-mail:\email{iandric@irb.hr},
\email{larisa@irb.hr},
\email{djurman@irb.hr}}

\abstract{
We analyse a specific, duality-based generalization of the 
hermitean matrix model. 
The existence of two collective fields enables us to describe 
specific excitations of the hermitean matrix model.
By using these two fields, we construct topologically non-trivial 
 solutions (BPS solitons)
 of the model. We find the 
low-energy spectrum of quantum fluctuations around the uniform solution.
Furthermore, we construct
the wave functional of the ground state and obtain the corresponding Green
function.}


\begin{document}

\section{Introduction}

The recently renewed interest in matrix models came from the 
identification~\cite{str}
of matrix quantum mechanics with the worldvolume theory of
unstable $D0$-branes in the two-dimensional string theory.
One uses matrix models to study non-perturbative aspects 
of string theory, or more specifically, the open-closed string duality.
 
In the usual approach, the closed string states are related to the matrix 
model states by bosonization of the fermion field followed by a non-local field 
redefinition~\cite{pol}. 
Alternatively, one can use the collective-field formulation of 
the matrix model to obtain a continuous, field-theoretic
description of the model.
This approach was used to construct 
the collective-field formulation of  unitary ~\cite{js} and 
symplectic~\cite{ajl} matrix models.
The 
important result of the  collective-field formulation was the analysis 
of higher-order terms in the $1/N$ expansion. It was shown that divergences 
cancelled and the finite correction, of order $1/N^2$, 
was calculated ~\cite{ab}.

The two-dimensional string theory is obtained from a matrix model
in the double scaling limit, with the critical potential given by the
inverted harmonic oscillator. It was shown in Ref.\cite{aj} that 
one could remove the oscillator potential from the
action by a suitable coordinate transformation. Therefore, it is of interest 
to study general features of the background independent
theory without the external potential.

Furthermore, it was realized that in order to describe black-hole solution, the
generalization of the standard hermitean matrix model is needed~\cite{ant}. 
One such generalization was proposed in the paper~\cite{5}. 

In this paper, we analyze  this duality-based generalization~\cite{5} 
and interpret 
the model in question as hermitean matrix model 
beyond the usual singlet approximation. 
We construct non-trivial solitonic solutions of the model,
and  find the spectrum of  quantum fluctuations around 
the uniform density solution of the master \h \ constructed in Refs.\cite{3,5}.

The master \h \ can be interpreted as a two-family Calogero model, connected
by duality, with mutually 
inverse coupling constants~\cite{3}, or, for a 
special choice of coupling strength ($\l=1/2$), as a master \h \ for
 symmetric and quaternionic models~\cite{5}. In Ref.\cite{5}
it was conjectured that the master \h \ with $\l=1/2$ actually 
corresponded to the hermitean matrix model, and our analysis of 
the model confirms this conjecture.
From the string point of view,
we describe oriented string degrees of freedom (hermitean matrix model)
by unoriented string degrees of freedom (symmetric and quaternionic 
matrix model)\footnote{One can think of our master \h \  as an 
explicit realization of an idea proposed in~\cite{4}, that one can 
continuously deform oriented string theories in two dimensions into 
unoriented ones by turning on non-local interactions on the worldsheet.}.
Although we focus on the matrix model with $\l=1/2$, constructed in 
Ref.\cite{5}, 
throughout the paper we write  $\l$ for generality. Also, in this
way, the  $\l\leftrightarrow 1/\l$ duality is obvious. 
The aim of this investigation is to shed more light on the nature of duality 
 in the field-theoretic 
description of  matrix 
models.  

\section{Semiclassical solutions of the master Hamiltonian}

Let us start with the master \h \ obtained in Ref.\cite{5}:
\baa\label{h1}
H^{M}&=&\frac{1}{2}\int dx \r(x)(\p_x\pi_{\r}(x))^2+
\frac{1}{8}\int dx\r(x)
\left[(\l-1)\frac{\p_x\r(x)}{\r(x)}+2\l\pv\int dy \frac{\r(y)}{x-y}+
2\pv\int dy\frac{m(y)}{x-y}\right]^2+\nn
&+&\frac{\l}{2}\int dx m(x)(\p_x\pi_{m}(x))^2+
\frac{\l}{8}\int dxm(x)
\left[(\frac{1}{\l}-1)\frac{\p_x m(x)}{m(x)}+\frac{2}{\l}\pv\int dy 
\frac{m(y)}{x-y}+
2\pv\int dy\frac{\r(y)}{x-y}\right]^2-\nn
&-&\frac{\l}{2}\int dx \left.\r(x)\p_x\frac{P}{x-y}\right|_{y=x}-
\frac{1}{2}\int dx \left.m(x)\p_x\frac{P}{x-y}\right|_{y=x}.\ea
The collective fields $\r(x)$ and $m(x)$ are normalized as
\baa\label{norm}
\int dx \r(x)=N,\;\int dx m(x)=M, \ea
where $N$ and  $M$ are  large numbers of independent matrix eigenvalues (or 
particles in the system for the Calogero model), 
 and $\pi_{\r/m}(x)$ represent the canonical momenta:
\baa\label{can}
\left[\p_x\pi_{\r}(x),\r(y)\right]=-i\p_x\delta(x-y),\;
\left[\p_x\pi_{m}(x),m(y)\right]=-i\p_x\delta(x-y).\ea
One can add a term $-\mu_{\r}\int dx \r(x)$ and an analogous one for $m(x)$
to the \h \ to ensure (\ref{norm}).

The leading part\footnote{Note that the divergent terms (the last line in 
the \h \ (\ref{h1}), as well as the momenta,  
are suppressed in the leading order, but are contributing to the spectrum of
 quantum  fluctuations.}
of the collective-field \h \ in the $1/N$ and 
$1/M$ expansions is 
given by the positive-definite effective potential, the second and 
the fourth term in the \h \ (\ref{h1}).
Owing to the positive definiteness of the effective potential,
the Bogomol'nyi limit appears. The Bogomol'nyi bound is saturated by the 
positive normalizable solutions $\rn(x)$ and $\m(x)$ of the coupled
equations
\baa\label{gs1}
&&(\l-1)\frac{\p_x\r(x)}{\r(x)}+
2\l\pv\int dy\frac{\r(y)}{x-y}+2\pv\int dy\frac{m(y)}{x-y}=0,\\
\label{gs2}
&&\left(\frac{1}{\l}-1\right)\frac{\p_x
m(x)}{m(x)}+\frac{2}{\l}\pv\int dy\frac{m(y)}{x-y}
+2\pv\int dy\frac{\r(y)}{x-y}=0.
\ea

An obvious solution is a constant density  for both fields: 
\baa\label{kond}
\rn=N/L,\;{\rm and} \;
\m=M/L.\ea
This solution exists on the compact support only, as can be seen from
the normalization conditions (\ref{norm}).

Next, let us look for non-trivial solutions of the coupled equations
(\ref{gs1}) and (\ref{gs2}).
Substracting these equations and then integrating them,  we find the condition
\baa\label{cc}
\rho(x)m(x)=c,
\ea
where c is an arbitrary constant parameter. Expressing $m(x)$ and introducing 
it in  eq. (\ref{gs1}),
we are left with
\begin{equation}
 (\lambda - 1) \partial_{x} \rho(x)+
  2\lambda \rho(x)\pv \int \frac{ dy \rho(y)}{x - y}  + 2c \rho(x)\pv \int \frac{dy }{ \rho(y) (x - y)} = 0.
  \end{equation}
Let us take the following ansatz
\begin{equation}
\rho(x) = \rho_{0} \frac { x^{2} + a^{2}}{x^{2} + b^{2}},
    \end{equation}
where $ \; a \; $ and $ \; b \; $ are arbitrary positive constants. 
Using the Hilbert transform
    \begin{equation}
     \pv \int \frac {dy}{x - y} \frac{1}{y-z} = i\pi \frac{sign(Im\;z)}{x-z},
     \end{equation}
     we find the conditions
     \baa
     \frac{\rho_{0} \pi }{b} ( b^{2} - a^{2}) = \frac{1-\lambda}{\lambda},\;
     c=\frac{\lambda \rn^2 a}{b}.
     \ea
The soliton-antisoliton  solution is given by
     \baa\label{rr1}
&&\rho(x) = \rho_{0} + \frac{\lambda - 1}{\lambda \pi} \frac{b}{x^{2} + b^{2}},
       \\
\label{rm1}
&&m(x) = \frac{c}{ \rho_{0}} + \frac{1 - \lambda}{ \pi} \frac{a}{x^{2} + a^{2}}.
\ea
Depending on the sign of $(\l-1)$, one density describes a hole 
in the condensate, whereas the other describes 
a particle above the condensate. 
We see that $\l\leftrightarrow 1/\l$ interchanges holes and particles. 

Now let us study the case $c=ka$ in the limit $a\rightarrow 0$.
	In this limit,  $m(x$) goes to the delta function:
	\baa\label{a1m}
	\lim_{a \rightarrow 0} m(x)= \frac{kb^2}{\rn}\pi\delta(x)
	\ea
	and $\rho(x)$ reduces to
	\baa\label{a1r}
	\lim_{a \rightarrow 0} \rho(x)=\rn \frac{x^2}{x^2+b^2}.
	\ea
Although these two expressions are obviously solutions 
of the Bogomol'nyi equations, it is
instructive to insert them into  eqs. (\ref{gs1}), (\ref{gs2}) to confirm 
the correctness of the limiting procedure.
Inserting (\ref{a1m}) and (\ref{a1r}) into 
eq. (\ref{gs1}), we obtain
the conditions
\baa\label{con1}
k=\frac{\lambda \rn^2}{b}\;\;,\;\pi \rn b=\frac{1-\lambda}{\lambda}.
\ea
Next, we check the validity of eq. (\ref{gs2}):
\baa\label{bogeq2del}
(1-\lambda)\partial_{x} m(x)
  +2 m(x) \pv \int dy \frac{m(y)}{x - y}
    +2 \lambda m(x) \pv \int dy \frac{\rho(y)}{x - y} =0.
    \ea

Inserting (\ref{a1m}) and (\ref{a1r}) into (\ref{bogeq2del}) we find
    \baa
    &&(1-\lambda)\lim_{a \rightarrow 0}\frac{ka}{\rn} \frac{2x(x^2+a^2)-2x(x^2+b^2)}{(x^2+a^2)^2}+
    2\lambda \lim_{a \rightarrow 0} \frac{ka}{\rn} \frac{x^2+b^2}{x^2+a^2}
     \lim_{a \rightarrow 0} \frac{\pi \rn (a^2-b^2)}{b}\frac{x}{x^2+b^2}+\nn
     &&+2 \lim_{a \rightarrow 0} \frac{ka}{\rn} \frac{x^2+b^2}{x^2+a^2}
     \lim_{a \rightarrow 0} \frac{\pi k (b^2-a^2)}{\rn}\frac{x}{x^2+a^2}=0.
     \ea
Using the identity
     \baa\label{i1}
     \lim_{b \rightarrow 0} \frac{2bx}{\left(x^2+b^2\right)^2}
     = 2\pi\frac{P}{x}\delta(x)=-\pi \p_x \delta(x)\ea
     and (\ref{a1m}), we obtain only one condition
     \baa
     \frac{\pi k b^2}{\rn}=1-\lambda
     \ea
which is obviously consistent with (\ref{con1}).
This is due to
$m(x)\sim \delta(x)$ and (\ref{bogeq2del}).
     
 Next, we construct the  two-soliton solution  of eqs. (\ref{gs1}), (\ref{gs2}).
 With the following ansatz for $\r(x)$ and $m(x)$
\baa\label{an1}
\r(x)&=&\rn \frac{(x-a)(x-a^*)(x+a)(x+a^*)}{(x-b)(x-b^*)(x+b)(x+b^*)},\nn
m(x)&=&\frac{c}{\r}=m_0 \frac {(x-b)(x-b^*)(x+b)(x+b^*)}{(x-a)(x-a^*)(x+a)
(x+a^*)},\ea
we easily find the condition $b=a$, so that the solutions reduce to trivial ones
with constant $\rn$ and $m_0$.

Now, let us take $c=kIm(a)$ in the limit $Im(a)\rightarrow0$, and 
write  the ansatz (\ref{an1}) in the more natural form
\baa\label{ansatzr}
\r(x)&=& \lim_{\epsilon \rightarrow 0}\rn\frac{[(x-x_0)^2+\epsilon^2]
[(x+x_0)^2+\epsilon^2]}{(x^2-b^2)(x^2-{b^*}^2)}=
\rn\frac{(x-x_0)^2(x+x_0)^2}{(x^2-b^2)(x^2-{b^*}^2)}=\nn
&=&\rn \left( 1+\frac{B}{x-b}+ \frac{B^*}{x-b^*}-\frac{B}{x+b}
-\frac{B^*}{x+b^*}\right)\;\;,\\
\label{ansatzm}
m(x)&=&\lim_{\epsilon \rightarrow 0} \frac{k\epsilon}{\rn} 
\frac {(x^2-b^2)(x^2-{b^*}^2)}{[(x-x_0)^2+\epsilon^2][(x+x_0)^2
+\epsilon^2]}=\tilde{k}\left[\delta(x-x_0)+\delta(x+x_0)\right],\ea
where
\baa\label{pokr}
B&=&\frac{(b^2-x_0^2)^2}{2b(b^2-{b^*}^2)},\nn
\tilde{k}&=&\frac {k\pi(x_0^2-b^2)(x_0^2-{b^*}^2)}{4\rn x_0^2}.
\ea
Inserting the ansatz (\ref{ansatzr})  into eq. 
(\ref{gs1}) we find the following three conditions:
     \baa\label{b1}
     B=-B^*\;\;,\;2\lambda \rn \pi i B=\lambda-1\;\;,\;\tilde{k}=1-\lambda.
     \ea
After we introduce $b=|b|e^{i\phi}$ and $r=x_0^2/{|b|^2}>0$,
the condition $B=-B^*$ gives
     \baa\label{b2}
     r=2\cos \phi -1 \Rightarrow x_0=|b|\sqrt{(2\cos \phi -1)},
     \ea
Next, from $\rn \pi i B=\lambda-1$ we find
     \baa
 \rn \pi |b|=\frac{1-\lambda}{\lambda} \frac{\cot{\frac{\phi}{2}}}{2\cos{\phi}}.
 \ea
So, from (\ref{b1}) and (\ref{b2}) we have obtained $|b|$ and $x_0$ 
 as functions of $\phi$ and
 the last condition gives $k$
 \baa
 \frac{2k\pi}{\rn}\left[ \frac {\cos^2{\phi} (1-\cos \phi)}{2\cos{\phi}-1}\right]=1-\lambda,
 \ea
 as a function of $\phi$.

In the BPS limit, one can construct $n$-soliton solutions, $n>2$,
following the strategy demostrated above for
the two-soliton solution. 
The general ansatz for $n$-soliton solution is of the form
\baa\label{opcin}
 m(z)&=&\sum_{i=1}^n m_0\delta(z-x_i),\nn
\r(x)&=&\rn\left(1+\sum_{\alpha=1}^n\left(\frac{B_{\alpha}}{x-z_{\alpha}}
+\frac{B^*_{\alpha}}{x-z^*_{\alpha}}\right)\right)\nn
&=&\rn\frac{\prod_{i=1}^n(x-x_i)^2}{\prod_{\alpha=1}^n(x-z_{\alpha})
(x-z^*_{\alpha})}.\ea
Note that this ansatz satisfies the condition $\r(x) m(x)=0$.
To obtain relations between poles and zeros
explicitly, one has  to solve the general algebraic 
problem of finding zeros of the polynomial of degree $n\geq 3$.
The soliton solutions have the following 
properties: they are located at the point at which the $\r(x)$ field
is vanishing, but the  $m(x)$ field  is becoming undetermined. 
This property is 
also characteristic of BPS solutions describing monopols and Julia-Zee dyons
\cite{JZ}.

\section{Quantum excitations}

\subsection{The spectrum of quantum fluctuations}

At  this point we analyse the dynamics of the collective-field 
excitations around the ground-state  solution (\ref{kond}) 
of our master \h \ (\ref{h1}).
First, we rewrite the master \h \ in the following form:
\baa\label{hsh}
H^{M}=\frac{1}{2}\int dx\r(x)A_0^{\dagger}(x)A_0(x)+
\frac{\l}{2}\int dxm(x)B_0^{\dagger}(x)B_0(x),\ea
where
\baa\label{opa}
&&A_0(x)=\p_x\pi_{\r}(x)+i\left[\frac{(\l-1)}{2}\frac{\p_x\r(x)}{\r(x)}+
\l\pv\int dy\frac{\r(y)}{x-y}+\pv\int dy\frac{m(y)}{x-y}\right],\nn
&&B_0(x)=\p_x\pi_m(x)+i\left[\frac{(1-\l)}{2\l}\frac{\p_x
m(x)}{m(x)}+\frac{1}{\l}\pv\int dy\frac{m(y)}{x-y}
+\pv\int dy\frac{\r(y)}{x-y}\right].
\ea
Next, we perform the $1/N$ ($1/M$) expansion of the collective field $\r (x)$  
($m(x)$) 
\baa\label{e1}
\r (x)=\rn+\p_x\e(x),\; m(x)=\m+\p_x\te(x), \ea
where $\p_x\e(x)$ and $\p_x\te(x)$ are small density quantum 
fluctations\footnote{The small density quantum fluctations are defined 
by the  explicit derivative because of the  normalization condition $\int dx 
\p_x \e(x)=0$.}.
We insert (\ref{e1}) into the \h \ (\ref{hsh}) and expand up to second-order 
terms in $\p_x\e(x)$ and $\p_x\te(x)$. We obtain a \h \ quadratic in 
fluctuations:
\baa\label{h22}
H^{(2)}=\frac{\rn}{2}\int dx A^{\dagger}A+\frac{\l\m}{2}\int dxB^{\dagger}B.
\ea
The operators $A(x)$ and $B(x)$
\baa\label{op}
A(x)&=&-\pi_\e(x)+i\left[\frac{(\l-1)}{2}\frac{\p_x^2\e}{\rn}
+\pv\int \frac{dy}{x-y}
(\l\p_y\e(y)+\p_y\te(y))\right],\\
B(x)&=&-\pi_{\te}(x)+i\left[\frac{(1-\l)}{2\l}\frac{\p_x^2\te}{\m}
+\pv\int \frac{dy}{x-y}
(\p_y\e(y)+\frac{1}{\l}\p_y\te(y))\right],
\ea
satisfy the following commutation relations:
\baa
\left[A(x),A^{\dagger}(y)\right]&=& -\frac{\l-1}{\rn}\p_x\p_y\delta(x-y)+2\l
\p_x\frac{P}{x-y},\label{comma} \\
\left[B(x),B^{\dagger}(y)\right]&=& -\frac{1-\l}{\l \m}\p_x\p_y\delta(x-y)+
 \frac{2}{\l}\p_x\frac{P}{x-y},\label{commb}\\
 \left[A(x),B^{\dagger}(y)\right]&=&2\p_x\frac{P}{x-y}\label{commab}.
 \ea

Now, to find the spectrum of low-lying excitations, we have to diagonalize the 
\h \ (\ref{h22}).
We expand  the operators $A(x)$ and $B(x)$ in terms of  new, complete 
sets of operators
\baa
A(x)&=&\sum_n\phi_n(x)a_n,\; A^{\dagger}(x)=\sum_n\phi_n^*(x)a_n^{\dagger},
\label{e2a}\\
B(x)&=&\sum_n\varphi_n(x)b_n,\; 
B^{\dagger}(x)=\sum_n\varphi_n^*(x)b_n^{\dagger},
\label{e2b}\ea
where
\baa\label{comm2}
\left[a_n,a^{\dagger}_m\right]=\o_n\delta_{nm},\;
\left[b_n,b^{\dagger}_m\right]=\O_n\delta_{nm},\;
\left[a_n,b_m^{\dagger}\right]\neq 0.\ea
We demand that the \h \ (\ref{h22}) should take the following form:
\baa\label{dh}
H^{(2)}=\sum_na^{\dagger}_na_n+\sum_nb^{\dagger}_nb_n.\ea
Generally, $n$ is the quantum number, 
and it is assumed that the sum is replaced 
by an integral  for a continuous spectrum. 

Let us concentrate on the first part of the 
\h \ (\ref{dh}). We insert the expansion (\ref{e2a}) in the 
commutators (\ref{comma}) and apply (\ref{comm2}) to obtain 
the completeness relation
\baa\label{compl}
\sum_n\o_n\phi_n(x)\phi_n^*(y)=-\frac{\l-1}{\rn}\p_x\p_y\delta(x-y)+2\l
\p_x\frac{P}{x-y}.\ea
Inserting (\ref{e2a}) in (\ref{h22}) and demanding (\ref{dh}), we obtain
an orthogonality relation
\baa \label{medju}
\frac{\rn}{2}\int dx \phi^*_n(x)\phi_m(x)=\delta_{nm}.\ea
Next, we multiply the relation (\ref{medju}) by $\phi^*_m(y)$, 
sum over $m$, apply
the completeness relation (\ref{compl}) and finally obtain the 
equation for the functions $\phi_n(x)$:
\baa\label{eq1}
\o_n\phi_n(x)=\frac{\l-1}{2}\p_x^2\phi_n(x)+\l\rn\p_x\pv\int dy\frac{\phi_n(y)}
{x-y}.\ea
The plane waves are the solution of Eq.(\ref{eq1}) with 
\baa\label{o}
\o_n=\frac{1-\l}{2} n^2+\l\rn\pi |n|.\ea
An analogous procedure for the second part of the \h \ (\ref{dh}) gives us 
$\O_n$
\baa\label{O}
\O_n=\frac{\l-1}{2}n^2+\m\pi |n|.\ea

A few comments are in order here. 
The relation (\ref{o}) was obtained in Ref.\cite{calog} as 
a dispersion relation for quantum fluctuations around the 
ground-state solution in the Calogero model. 
Under the duality transformation $\l\rightarrow 1/\l$ and 
$\rn\leftrightarrow\m$, the relations (\ref{o}) and (\ref{O}) are interchanged, 
reflecting the known duality symmetry  
$\l\rightarrow 1/\l$  of the Calogero model.
However, note that these dispersion relations  do not describe physical 
excitations since $[a,b^{\dagger}]\neq 0$, i.e., the \h \ (\ref{dh}) is not
diagonal!

Furthermore,
for $\l=1/2$,
the  relations (\ref{o}) and (\ref{O}) 
were obtained 
in Ref.\cite{ap}. In this paper, these dispersions represented 
quantum fluctuations for two independent, effective \h s operating 
on two particular subsets of eigenstates of the hermitean matrix model\footnote{
These effective \h s were actually a collective-field formulation
of matrix models for the symmetric and 
quaternionic matrices, compare Ref.\cite{5}.}.
However, our duality-based generalization of the hermitean matrix model 
also includes the
interaction between these two sets of mutually dual degrees of freedom. 

Let us continue with our diagonalization procedure.
From the completeness and orthogonality relations we obtain
\baa\label{pf}
\phi_n(x)=\frac{\exp(inx)}{\sqrt{\pi\rn}},\;
\varphi_n(x)=\frac{\exp(inx)}{\sqrt{\pi\l\m}}.\ea
Now, we are in a position to calculate the commutator 
$\left[a_n,b_m^{\dagger}\right]$. From (\ref{commab}) and (\ref{medju}),
using (\ref{o}), (\ref{O}), and (\ref{pf}),
we obtain
\baa\label{comab}
\left[a_n,b_m^{\dagger}\right]=\delta_{nm}|n|\pi\sqrt{\rn\m\l}\equiv\delta_{nm}
f_n.\ea

In the limit of small $n$, an interesting
possibility  arises - $a_n\propto b_n$.
Appropriate rescaling  of the relations (\ref{comm2})  gives us
the following \h :
\baa\label{small}
H^{(2)}=\sum_n(\o_n+\O_n)c_n^{\dagger}c_n=\sum_n |n|\pi(\l\rn+\m)
c_n^{\dagger}c_n, \ea
where the  bosonic operators $c_n$ satisfy
\baa\label{comc}
\left[c_n,c_m\right]=[c_n^{\dagger},c_m^{\dagger}]=0,\;
\left[c_n,c_m^{\dagger}\right]=\delta_{nm}
\ea
We see  that in the 
lowest order, the  special case $\l=1/2$ reproduces the 
part of singlet sector of hermitean matrix model~\cite{calog}, i.e., 
we can represent the hermitean matrix model 
(the Calogero model with $\l=1$) as 
a dual system of symmetric ($\l=1/2$) and quaternionic ($\l=2$) matrices.

Of course, this is to be expected. Although we claim that the collective-field variables
used in the construction ~\cite{5} of the master \h \ (\ref{hsh}) with $\l=1/2$
do contain some
non-singlet states of the hermitean matrix 
model,  these non-singlet states
do not contribute in the low-energy approximation, and, therefore, the obtained
result is in fact  a consistency check.
In the string theory language, this means that the unoriented string 
amplitudes at tree level are the same as in the oriented model.

What can we say about 
the excitations for "not-so-low" energy?
Let us exactly diagonalize the \h \ (\ref{dh}). 
Having the quadratic  \h \ allows us to apply the Bogoliubov transformation.
We introduce two sets of mutually commuting  operators $\tc_n$ and 
$\td_n$ 
\baa\label{comcd}
\left[\tc_n,\tc^{\dagger}_m\right]=\o_n^+\delta_{nm},\;
\left[\td_n,\td^{\dagger}_m\right]=\o_n^-\delta_{nm},\ea
such that we can write the \h \ (\ref{dh}) as a sum of two independent 
quadratic parts:
\baa\label{dhf}
H^{(2)}=\sum_n\tc^{\dagger}_n\tc_n+\sum_n\td^{\dagger}_n\td_n.\ea
Assume that 
\baa\label{b}
a_n=\alpha_n \tc_n-\beta_n \td_n\; {\rm and}\; b_n=\beta_n \tc_n
+\alpha_n \td_n, \ea
where $\alpha_n,\beta_n$ are some 
coefficients that depend on the quantum number $n$.
Now, inserting (\ref{b}) into (\ref{dh}) and demanding (\ref{dhf}), and 
inserting (\ref{b}) into (\ref{comm2}), gives us a system of four equations
\baa\label{sys}
\alpha_n\beta_n(\o_n^+-\o_n^-)=f_n,\;\alpha^2_n+\beta^2_n=1,\nn
\alpha_n^2\o_n^++\beta^2_n\o_n^-=\o_n ,\;
\beta^2_n\o_n^++\alpha^2_n\o_n^-=\O_n,
\ea
which we solve.
The coefficients in the Bogoliubov transformation (\ref{b}) are
\baa\label{bc}
\alpha^2_n&=&\frac{1}{2}\left(1+\frac{\o_n-\O_n}{\sqrt{(\o_n-\O_n)^2+
4f_n^2}}\right),\nn
\beta^2_n&=&\frac{1}{2}\left(1-\frac{\o_n-\O_n}{\sqrt{(\o_n-\O_n)^2+
4f_n^2}}\right)
.\ea
The spectrum of low-lying excitations is given by
\baa\label{o1}
\o_n^{\pm}&=&\frac{1}{2}\left[ \o_n+\O_n\pm\sqrt{(\o_n-\O_n)^2+
4f_n^2}\right]=\nn
&=&\frac{\pi(\l\rn+\m)|n|}{2}\left(1\pm\sqrt{1+\frac{2(1-\l)(\l\rn-\m)}{\pi
(\l\rn+m)^2}|n|+\frac{(1-\l)^2}{\pi^2(\l\rn+\m)^2}n^2}\right).\ea
Note  that each branch of the dispersion relation (\ref{o1}) is invariant 
under the 
duality transformation $\l\rightarrow 1/\l$ and  $\rn\leftrightarrow\m$,
and physical excitations are not connected by duality. We have two 
sets of independent oscillators.
Also note that  obtained corrections are of the same order as non-singlet 
corrections~\cite{kl}, namely $1/N^2$ with respect to 
the leading singlet contribution.

However, for $\l=1/2$ the ground system is  dual, i.e.,  $\l\rn=\m$
and $\o_n^-$ is negative!
The "wrong" sign of $\o_n^-$ signalizes the instability of the vacuum.

\subsection{The vacuum functional and Green functions}

The vacuum functional is a simultaneous solution of the equations
\baa\label{fun}
A_0(x)|0\rangle=0\;{\rm and}\; B_0(x)|0\rangle=0,\ea
and is given by
\baa\label{vac}
\Psi_0(\r,m)&=&\exp\left\{\frac{(\l-1)}{2}\int dx \r(x)\ln\r(x)+\frac{\l}{2}
\int\int dx dy \r(x)\ln|x-y|\r(y)\right.
+\nn &+&\int\int dx dy\r(x)\ln|x-y|m(y)+\nn
&+&\left.
\frac{(1-\l)}{2\l}\int dx m(x)\ln m(y)+\frac{1}{2\l}
\int\int dx dym(x)\ln|x-y|m(y)\right\}.\ea
We expand the vacuum found in (\ref{vac}) up 
to quadratic terms
\baa\label{vac2}
\Psi^{(2)}=\exp \left( \int\int dzdy\; \boldsymbol{\eta}^{T}(z)
{\bf G}^{-1}(z-y) \boldsymbol{\eta}(y) \right), 
\ea
where we have introduced the matrix notation
\baa\label{prop}
\boldsymbol{\eta}^T(y)&=&
\left[
\frac{\p_y\eta(y)}{\sqrt{\rho_0}}\;,\;
\frac{\p_y\tilde{\eta}(y)}{\sqrt{\lambda m_0}}
\right]\nn
{\bf G}^{-1}(z-y)&=&-\frac{1}{4\pi}
\left[\begin{smallmatrix}
(1-\l)\pi \delta(z-y)-2\l \pi \rho_0 \ln |z-y|& -2 \pi \sqrt{\lambda \rho_0 m_0} \ln|z-y|\\
 -2 \pi \sqrt{\lambda \rho_0 m_0} \ln|z-y|& (\l-1) \pi \delta(z-y)-2 \pi m_0 \ln |z-y|
\end{smallmatrix}\right]=\nn
&=&-\frac{1}{4\pi}\int dk \frac{e^{ik(z-y)}}{k^2}
\left[\begin{smallmatrix}
\omega_k & f_k\\
f_k & \Omega_k
\end{smallmatrix}\right]\;.
\ea
This vacuum is a solution of the  equations
\baa\label{vaceq}
A(x) \Psi^{(2)}=0, B(x) \Psi^{(2)}=0 \Leftrightarrow \tilde{c}_k \Psi^{(2)}=0, \tilde{d}_k \Psi^{(2)}=0
\ea
Writing $\Psi^{(2)}$ in terms of the field 
\baa
\boldsymbol{\varphi}(x)&=&\int dx {\bf {S}}(x-x')
\boldsymbol{\eta}(x'),\nn 
{\bf {S}}(x-x')&=&\int dk \frac{e^{ik(x-x')}}{2\pi\sqrt{|k|}(k^2+k_0^2)^{1/4}}
\left[\begin{smallmatrix}
\alpha_k & \beta_k\\
-\beta_k & \alpha_k
\end{smallmatrix}\right]\;,
\ea
where $k_0^2=4\l^2\pi^2\rn^2/(1-\l)^2$, we obtain
\baa
\Psi^{(2)}=\exp\left\{\int \int dzdy\boldsymbol{\varphi}^{T}(z)
\left(
-\frac{1}{4\pi}\int dk e^{ik(z-y)}\frac{\sqrt{k^2+k_0^2}}{|k|}
\left[\begin{smallmatrix}
\omega_{k}^{+} & 0\\
0 & \omega_{k}^{-}
\end{smallmatrix}\right]\right)
\boldsymbol{\varphi}(y)
\right\}
\ea
and the  non-normalizability of the vacuum $\Psi^{(2)}$ is manifest.
This is a consequence of the commutation relations (\ref{comcd})
from which we conclude that $\td_n$ is actually a creation operator
which we  call $d^{\dagger}_n$ and $\td^{\dagger}_n$ is 
an annihilation operator now called $d_n$.

The \h \ (\ref{dhf}) is, after appropriate rescaling, given by
\baa\label{dhf2}
H^{(2)}=\sum_n \o_n^+c^{\dagger}_nc_n+\sum_n|\o_n^-|d_n 
d^{\dagger}_n=\sum_n \o_n^+c^{\dagger}_nc_n+\sum_n|\o_n^-|
d_n^{\dagger}d_n+\sum_n|\o_n^-|.
\ea
The last term must be included in the vacuum energy. It simply
defines the Fermi level of the system.

Solving equations for the normalizable vacuum functional
\baa
c_k \Phi^{(2)}=0, d_k \Phi^{(2)}=0,
\ea
we find 
\baa\label{ndvc}
\Phi^{(2)}=\exp\left\{\int \int dzdy\boldsymbol{\varphi}^{T}(z)
{\bf G}^{-1}_{\varphi\varphi}(z-y)
\boldsymbol{\varphi}(y)
\right\}.
\ea
The inverse propagator  ${\bf G}^{-1}_{\varphi\varphi}(z-y)$ is given by
\baa\label{pp}
{\bf G}^{-1}_{\varphi\varphi}(z-y)&=&
-\frac{1}{4}\int dk e^{ik(z-y)}\frac{\sqrt{k^2+k_0^2}}{|k|}
\left[\begin{smallmatrix}
\omega_{k}^{+} & 0\\
0 & |\omega_{k}^{-}|
\end{smallmatrix}\right]=\nn
&=& \frac{\l-1}{4}(\p_z\p_y+k_0^2)\left[
\left(\begin{smallmatrix}
1 & 0 \\
0 & 1 \end{smallmatrix}\right)\delta(z-y)+
\left(\begin{smallmatrix}
1 & 0 \\
0 & -1 \end{smallmatrix}\right)\frac{k_0}{\pi}K_0(k_0(z-y))\right],\ea
where $K_0(x)$ is the Bessel function.

The two components of the field $\boldsymbol{\varphi}(y)$ can be combined into
a complex scalar field, which could be interpreted as
a complex tachyon field of 
$0A$ two-dimensional string theory~\cite{toum}.

 \section{Conclusion}

We have analysed the spectrum of  quantum fluctuations around a
particular ground-state solution of the master \h \ (\ref{hsh}),
representing a duality-based generalization of the hermitean  matrix model.
This has  given us
a glimpse of the  dynamics of  dual degrees of freedom in  field theory.
 We have shown that for $\l=1/2$, the master \h \ describes a new
representation of the hermitean matrix model, in which  some specific 
states  have been  analysed.
 These states contribute to the next-to-leading order, and the contribution is
 comparable with the contribution from the non-singlet sector.
Also, we are able to
 construct topologically non-trivial, solitonic solutions 
 of the model in question, which do not exist in the singlet approximation 
 of the hermitean matrix model. 
In Ref.\cite{kl} the non-singlet states were attributed to the vortices
on the worldsheet. 
Note  that these non-local interactions on the worldsheet
can be interpreted as double-trace deformations of the standard hermitean
matrix model originating from the $(\l-1)$-proportional terms
in the master \h ~\cite{4}.

The excitations in our model
 behave as quasi-particles and quasi-holes, and  can be described by a
 complex scalar field.
 Therefore, in our opinion, the relation between the matrix model
  constructed in Ref.\cite{5} and 0A string theory in two dimensions deserves
  a more detailed analysis.
 Furthermore,  we have constructed the
 soliton-antisoliton solution for the master \h , and  quantum fluctuations
 around this solution are currently  analysed~\cite{zad}. 
 Also, it would be interesting to investigate the possible
 integrability of the model, 
 as the
 existence of multisoliton solutions suggests.
   
\acknowledgments
This work was supported by the Ministry of Science and Technology of the
Republic of Croatia under contract  No. 0098003.
L. J.  gratefully acknowledges the
 partial support
 of
 the Alexander
 von Humboldt Foundation.


\begin{thebibliography}{99}
\bibitem{str}
J.~McGreevy and H.~Verlinde,
JHEP {\bf 0312} (2003) 054; I.~R.~Klebanov, J.~Maldacena, and N.~Seiberg,
JHEP {\bf 0307} (2003) 045
\bibitem{pol}
M. Natsuume and J. Polchinski, Nucl.\ Phys.\ B {\bf 424} (1994) 137.  
\bibitem{js}
A. Jevicki and B. Sakita, Nucl.\ Phys.\ B {\bf 165} (1980) 511.
\bibitem{ajl}
I. Andri\'c, A. Jevicki, and H. Levine, Nucl.\ Phys.\ B {\bf 215} (1983) 307.
\bibitem{ab}
I. Andri\'c and V. Bardek, Phys.\ Rev.\ D {\bf 32} (1985) 1025;
I. Andri\'c and V. Bardek, J.\ Phys.\ A {\bf 24} (1991) 353.
\bibitem{aj}
J. Avan and A. Jevicki, Phys.\ Lett.\ B {\bf 266} (1991) 35.
\bibitem{3}
I. Andri\'c and L. Jonke, Phys. Rev. A {\bf 65} (2002) 034707; 
I. Andri\'c and D. Jurman, Fortsch.\ Phys.\  {\bf 50} (2002) 675, 
Phys. Lett. A {\bf 313} (2003) 252.
\bibitem{ant}
A.~Jevicki and T.~Yoneya,
Nucl.\ Phys.\ B {\bf 411} (1994) 64;
U.~H.~Danielsson, J.~P.~Gregory, M.~E.~Olsson, P.~Rajan and M.~Vonk,
JHEP {\bf 0404} (2004) 065.
\bibitem{5}
I. Andri\'c and D. Jurman, 
JHEP {\bf 0501} (2005) 039.
\bibitem{4}
J. Gomis and A. Kapustin, JHEP {\bf 0406} (2004) 002.
\bibitem{JZ}
B. Julia and A. Zee, 
  Phys.\ Rev.\ D {\bf 11} (1975) 2227;
R. Teh, {\it A note on the Julia-Zee dyon solution},
    arXiv:hep-th/0505058.
\bibitem{calog}
I.~Andri\'c, V.~Bardek, and L.~Jonke,
Fizika {\bf B4} (1995) 93.
\bibitem{ap}
A. Jevicki, Nucl.\ Phys.\ B {\bf 376} (1992) 75.
\bibitem{toum}
T.~Takayanagi and N.~Toumbas,
JHEP {\bf 0307} (2003) 064 ; 
M.~R.~Douglas et. al., 
{\it A new hat for the c = 1 matrix model},
arXiv:hep-th/0307195.
\bibitem{kl}
D. J. Gross and I. Klebanov, 
Nucl.\ Phys.\ B {\bf 354} (1991) 459.

\bibitem{Jevicki:1979mb}
A.~Jevicki and B.~Sakita,
Nucl.\ Phys.\ B {\bf 165} (1980) 511;
A.~Jevicki and B.~Sakita,
Nucl.\ Phys.\ B {\bf 185} (1981) 89,
B.~Sakita, {\em Quantum Theory of 
Many Variable Systems and Fields} (World Scientific, 1988).
\bibitem{Andric:1994nc}
I.~Andri\'c, V.~Bardek and L.~Jonke,
Phys.\ Lett.\ B {\bf 357} (1995) 374.

\bibitem{Polychronakos:1994xg}
A.~P.~Polychronakos,
Phys.\ Rev.\ Lett.\  {\bf 74} (1995) 5153.

\bibitem{zad}
I.~Andri\'c, L.~Jonke and D.~Jurman, work in progress.
\end{thebibliography}
\end{document}